\documentclass{aa}

\usepackage{amsmath}
\usepackage{ulem}
\usepackage[varg]{txfonts}
\usepackage{natbib}
\bibpunct{(}{)}{;}{a}{}{,}

\usepackage{xspace}

\usepackage{hyperref}
\hypersetup{colorlinks=true, linkcolor=red,citecolor=cyan,filecolor=green,urlcolor=magenta}

\newcommand{\hl}[1]{\textcolor{brown}{\textbf{#1}}}
\renewcommand{\hl}[1]{#1}

\newcommand{\added}[1]{\textcolor{black}{\textbf{#1}}}
\renewcommand{\added}[1]{#1}

\newcommand{\removed}[1]{\sout{\textcolor{black}{\textbf{#1}}}}
\renewcommand{\removed}[1]{\ignorespaces}

\newcommand{\nnhp}{\mbox{N$_2$H$^+$}\xspace}

\newcommand{\nhhh}{\mbox{NH$_3$}\xspace}

\newcommand{\um}{\mbox{$\mu$m}\xspace}

\newcommand{\signt}{\mbox{$\sigma_{\rm nt}$}\xspace}

\newcommand{\kms}{\mbox{$\mathrm{km~s^{-1}}$}\xspace}

\newcommand{\Ms}{\mbox{$M_{\sun}$}\xspace}

\begin{document}

\title{Subsonic islands within a high-mass star-forming IRDC}

\titlerunning{Subsonic islands in IRDC G035.39}
\authorrunning{Sokolov et al.}

% TODO: ORCID identifier? [0000-0002-5327-4289] Does A&A support this?
\author{
         Vlas Sokolov\inst{\ref{mpe}}
    \and Ke Wang\inst{\ref{eso}}
    \and Jaime E. Pineda\inst{\ref{mpe}}
    \and Paola Caselli\inst{\ref{mpe}}
    \and Jonathan D. Henshaw\inst{\ref{mpia}}
    \and Ashley T. Barnes\inst{\ref{mpe}, \ref{ljmu}}
    \and Jonathan C. Tan\inst{\ref{florida-astro}, \ref{florida-ph}}
    \and Francesco Fontani\inst{\ref{arcetri}}
    \and Izaskun Jim\'enez-Serra\inst{\ref{queen-mary}}
    \and Qizhou Zhang\inst{\ref{cfa}}
}

\institute{
         Max Planck Institute for Extraterrestrial Physics, Gie{\ss}enbachstra{\ss}se 1, D-85748 Garching bei M{\"u}nchen, Germany \\ \email{vsokolov@mpe.mpg.de}\label{mpe}
    \and European Southern Observatory, Karl-Schwarzschild-Str. 2, D-85748, Garching bei M{\"u}nchen, Germany\label{eso}
    \and Max Planck Institute for Astronomy, K\"{o}nigstuhl 17, D-69117 Heidelberg, Germany\label{mpia}
    \and Astrophysics Research Institute, Liverpool John Moores University, 146 Brownlow Hill, Liverpool L3 5RF, UK\label{ljmu}
    \and Department of Astronomy, University of Florida, Gainesville, FL, 32611, USA\label{florida-astro}
    \and Department of Physics, University of Florida, Gainesville, FL, 32611, USA\label{florida-ph}
    \and INAF-Osservatorio Astrofisico di Arcetri, Largo E. Fermi 5, I-50125 Firenze, Italy\label{arcetri}
    \and School of Physics and Astronomy, Queen Mary University of London, Mile End Road, London E1 4NS, UK\label{queen-mary}
    \and Harvard-Smithsonian Center for Astrophysics, 60 Garden Street, Cambridge MA 02138, USA\label{cfa}
}

\date{Received 1 February 2018 / Accepted 19 February 2018}

\abstract{
% TODO: does A&A have word limit on the abstract?
High-mass star forming regions are typically thought to be dominated by supersonic motions. We present combined Very Large Array and Green Bank Telescope (VLA+GBT) observations
of \mbox{NH$_3$}\xspace (1,1) and (2,2)
in the infrared dark cloud (IRDC) G035.39-00.33, tracing cold and dense gas down to scales of 0.07 pc.
We find that, in contrast to previous, similar studies of IRDCs,
more than a third of the fitted ammonia spectra show subsonic non-thermal motions (mean line width of 0.71 \kms),
and the sonic Mach number distribution peaks around $\mathcal{M} = 1$.
As possible observational and instrumental biases would only broaden the line profiles, our results provide strong upper limits to the actual value of $\mathcal{M}$, further strengthening our findings of narrow line widths.
This finding calls for a reevaluation of the role of turbulent dissipation and subsonic regions in massive-star and cluster formation.
Based on our findings in G035.39, we further speculate that the coarser spectral resolution used in the previous
VLA \mbox{NH$_3$}\xspace studies
may have inhibited the detection of subsonic turbulence in IRDCs. The reduced turbulent support suggests that dynamically important magnetic fields of the 1 mG order would be required to support against possible gravitational collapse. Our results offer valuable input into the theories and simulations that aim to recreate the initial conditions of high-mass star and cluster formation.
}

\maketitle

\section{Introduction} \label{sec:intro}

Supersonic turbulence has been found to dominate molecular cloud kinematics in star forming regions \citep{larson1981}.
Towards the densest cores within these regions, however, the turbulence dissipates, allowing for the formation of stars \citep{benson+1989, foster+2009}.
Observationally, this decay of turbulence has been found to manifest itself as a difference in line widths between the lower-density gas, where the turbulence is scale-dependent, and the higher-density gas tracers, sensitive to the star-forming cores with typical sizes of 0.1 pc
%, revealing subsonic gas motions that are roughly constant in magnitude 
\citep{goodman+1998, caselli+2002}.
This change in gas kinematics, or transition to coherence, has been observationally identified to be a sharp boundary between subsonic and transonic motion regimes in the Perseus B5 region \citep{pineda+2010}.

While a number of studies have focused on identifying the subsonic motions in low-mass star-forming cores, massive star formation is typically thought to be accompanied by large non-thermal gas motions, as it usually proceeds in highly dynamical environments where the protostars, having shorter dynamical timescales, rapidly accrete material from their surroundings \citep[e.g.][]{tan+2014}.
Indeed, recent observations
point to the existence of large, pc-scale gas flows towards the star-forming cores \citep{peretto+2013, liu+2012a, liu+2012b, sanchez-monge+2013, henshaw+2014, liu+2015, wyrowski+2016}.
Moreover, non-thermal motions are an important element in the turbulent core model for massive star formation \citep{mckee+tan2003}, where turbulent motions provide extra support against the gravitational collapse, allowing for greater accretion rates and, ultimately, for a higher stellar mass.

A massive filamentary infrared dark cloud (IRDC)
%A massive filamentary IRDC
\object{G035.39-00.33} (G035.39 henceforth) at a distance of 2.9 kpc \citep{simon+2006} has been extensively studied in the past with a variety of continuum and molecular tracers. The non-thermal velocity dispersions inferred in G035.39 from the single-dish CO observations have been found to be two to three times higher than the sound speed in the medium \citep{jimenez-serra+2014}, and \cite{henshaw+2014} identified the filaments within the IRDC to be only mildly supersonic ($\mathcal{M} \sim 1.4-1.6$) with the 4\arcsec~Plateau de Bure Interferometer (PdBI) \nnhp (1-0) observations. Furthermore, on large, parsec scales, the molecular gas that forms the IRDC is not affected by the feedback from embedded protostars: widespread CO depletion \citep{hernandez+2011, hernandez+2012, jimenez-serra+2014}, as high deuterium fractionation levels \citep{barnes+2016} and low gas kinetic temperatures \citep{sokolov+2017} were found across the cloud. Despite the cloud's starless appearance at parsec scales, some star formation in the cloud is already underway (\cite{nguyen_luong+2011} find a number of compact 70\um sources in G035.39, some capable of forming intermediate- to high-mass stars).

By making use of the simultaneously derived temperature structure and the kinematics information obtained from the Very Large Array (VLA) on scales down to 0.07 pc, we are in a unique position to analyse dynamics of dense gas linking the cloud scales to those of the embedded protostars and protoclusters.
In this letter, we report the presence of widespread subsonic gas motions throughout G035.39.

\begin{figure}
    \centering
    \includegraphics[height=0.5\textwidth]{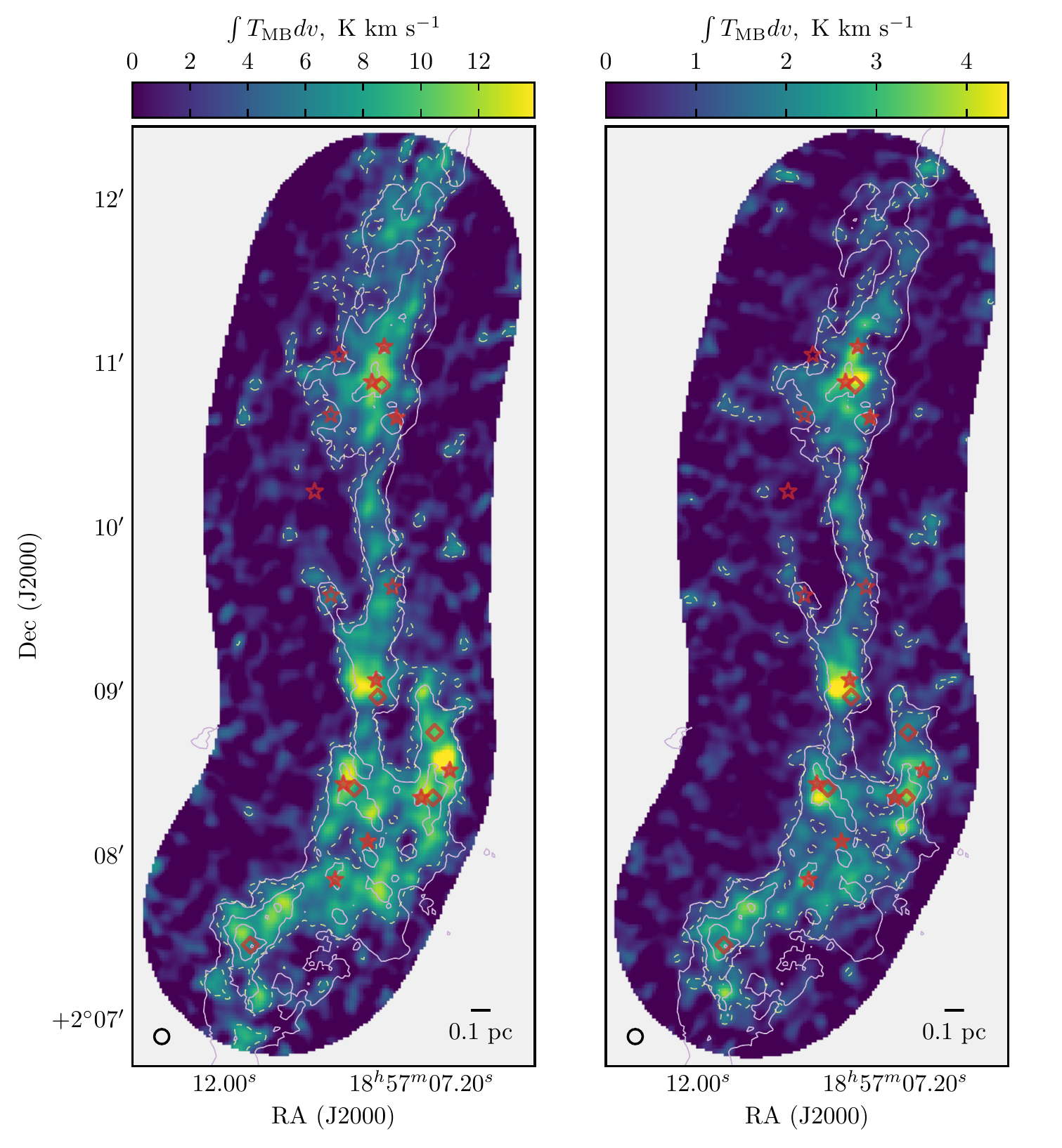}
    \caption{Left to right: \added{combined VLA+GBT integrated intensities} of the observed ammonia (1,1) and (2,2) lines, computed between 42 and 47 \kms. The white dotted contour marks the significance level of the integrated intensity of each line at the $3\sigma_I$ level, where $\sigma_I$ is the integrated intensity uncertainty \citep{mangum+shirley2015}. The white solid contours show the infrared extinction contours \citep{kainulainen+tan2013} starting from $A_{\rm V}=30$ mag and progressing inwards in steps of 30 mag. The open and filled red stars denote the positions of the Herschel sources from \cite{nguyen_luong+2011} below and above 20 \Ms, respectively, while the open diamonds mark the location of cores from \cite{butler+tan2009, butler+tan2012}.}
    \label{fig:overview}
\end{figure}

\section{Data reduction}

The VLA observations were conducted on May 8, 2010 (project AW776; PI: Ke Wang), in two consecutive sessions, mapping the \nhhh (1,1) and (2,2) inversion transitions in the compact D-configuration as a five-point mosaic covering the entire IRDC. The separate sessions allowed the spectral setup to achieve the complete hyperfine structure coverage of the \nhhh transitions with a resolution of 15.625 kHz.
The data were calibrated on the quasars J1851+005 (gain), J2253+1608 (bandpass), and 3C48 (flux) within the CASA data reduction package.

% NOTE: I got some neat weather summary images generated with the VLA pipeline ran on the two ammonia sessions. They are hiding under the '~/Data/g35.39/vla-pipeline/nh3[11|22]_kwang.ms.plotweather.png' globber. Wind speeds, temperatures, and humidities are easily readable there.
The calibrated visibilities were deconvolved with CASA task \textsc{tclean} using the multi-scale CLEAN algorithm \citep{cornwell2008}, with Briggs weighting and robust parameter set to 0.5. We taper the visibilities to achieve a similar synthesized beam between the two ammonia lines.
To account for the missing flux,
we fill in the zero spacing information from the ammonia data taken with the Green Bank Telescope (GBT). Detailed description of the GBT data reduction can be found in \cite{sokolov+2017}.
The GBT data were converted to spectral flux density units and smoothed to the VLA spectral resolution of 0.2 \kms with a Gaussian kernel. We then regrid both GBT (1,1) and (2,2) spectral cubes to the VLA spatial grid and apply the VLA mosaic primary beam response to the GBT images.
\hl{The VLA spectral cubes for both inversion lines are then deconvolved again with \textsc{tclean}, with the \textsc{tclean} mask being determined by combining the VLA only and GBT datasets using the \textsc{feather} task in CASA.
By constructing the clean mask from independently feathered images, we ensure that unbiased knowledge of the extended emission is incorporated into the \textsc{tclean} run.}
The resulting \nhhh (1,1) and (2,2) spectral cubes, gridded into 1\arcsec~pixels, have a common restoring beam of 5.44\arcsec.
The typical rms value of the emission free spectra in the resultant cubes is 14 mJy/beam for \nhhh (1,1) and 5 mJy/beam for \nhhh (2,2) inversion lines in a 0.2 \kms channel.

\section{Results} \label{sec:results}

Figure \ref{fig:overview} presents the integrated intensities of the \nhhh (1,1) and (2,2) lines overlaid with the mid-infrared extinction contours \citep{kainulainen+tan2013}.
The good correspondence between the two morphologies indicates that the dense gas, forming the bulk of the IRDC, is well traced with our \added{combined} ammonia observations. In addition, both inversion transitions have similar dynamic range (cf. 3$\sigma_I$ detection contours on Fig. \ref{fig:overview}), allowing us to reliably constrain the gas temperature that depends on the ratio of the two metastable inversion level populations \citep{ho+townes1983}.

\begin{figure*}
    \centering
    \includegraphics[height=8.5cm]{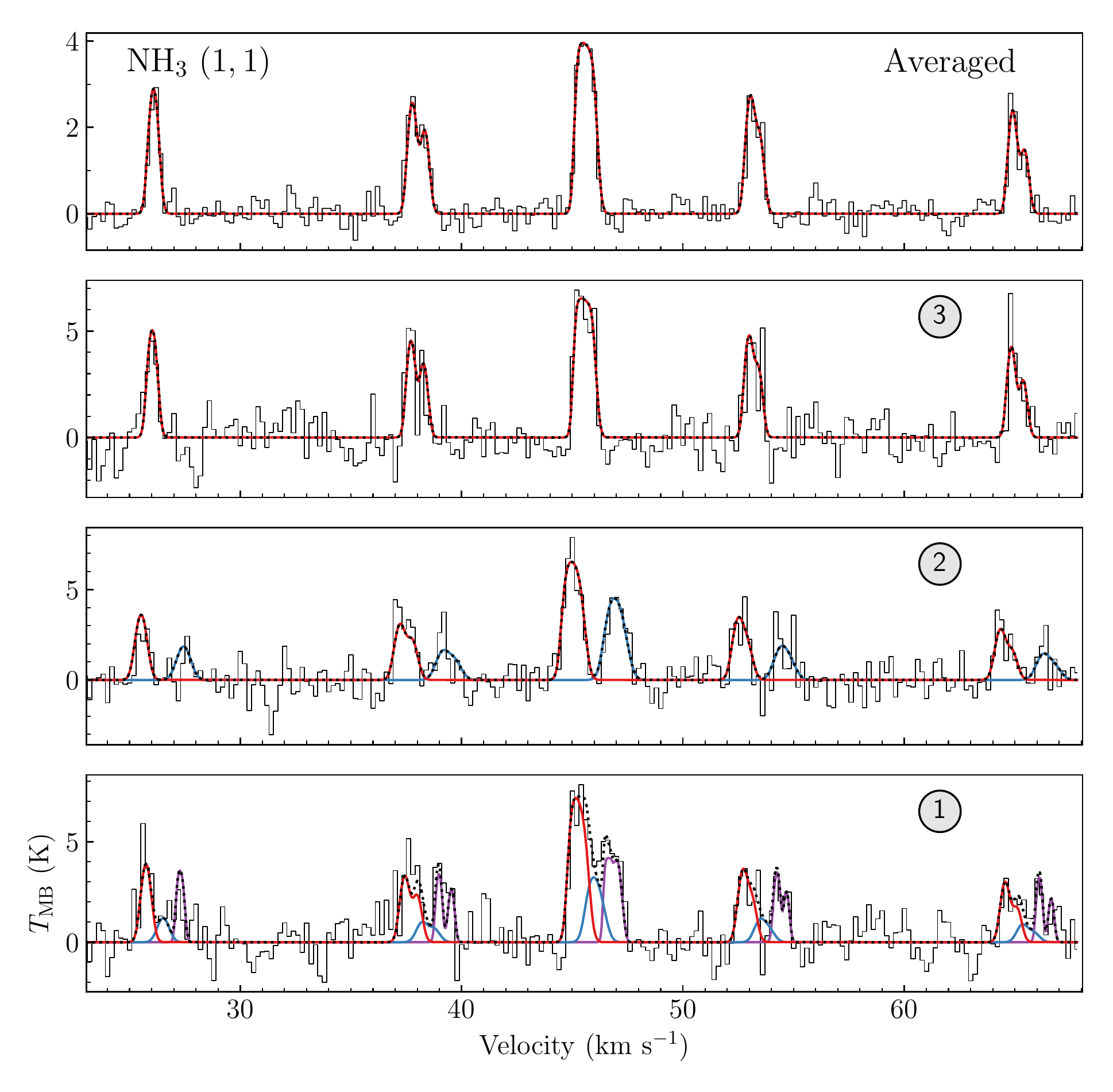}
    \includegraphics[height=8.5cm]{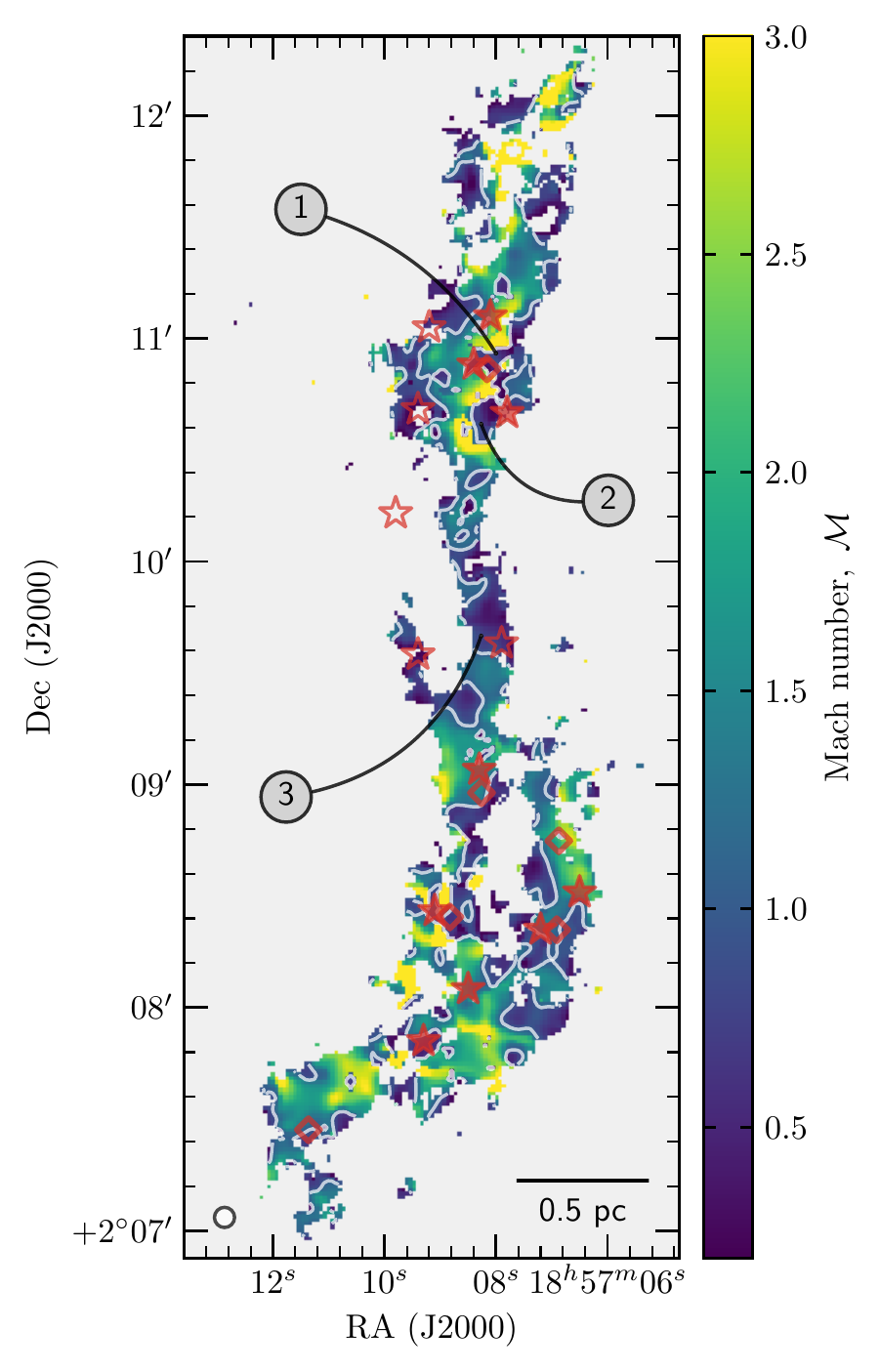}
    \includegraphics[height=8.48cm]{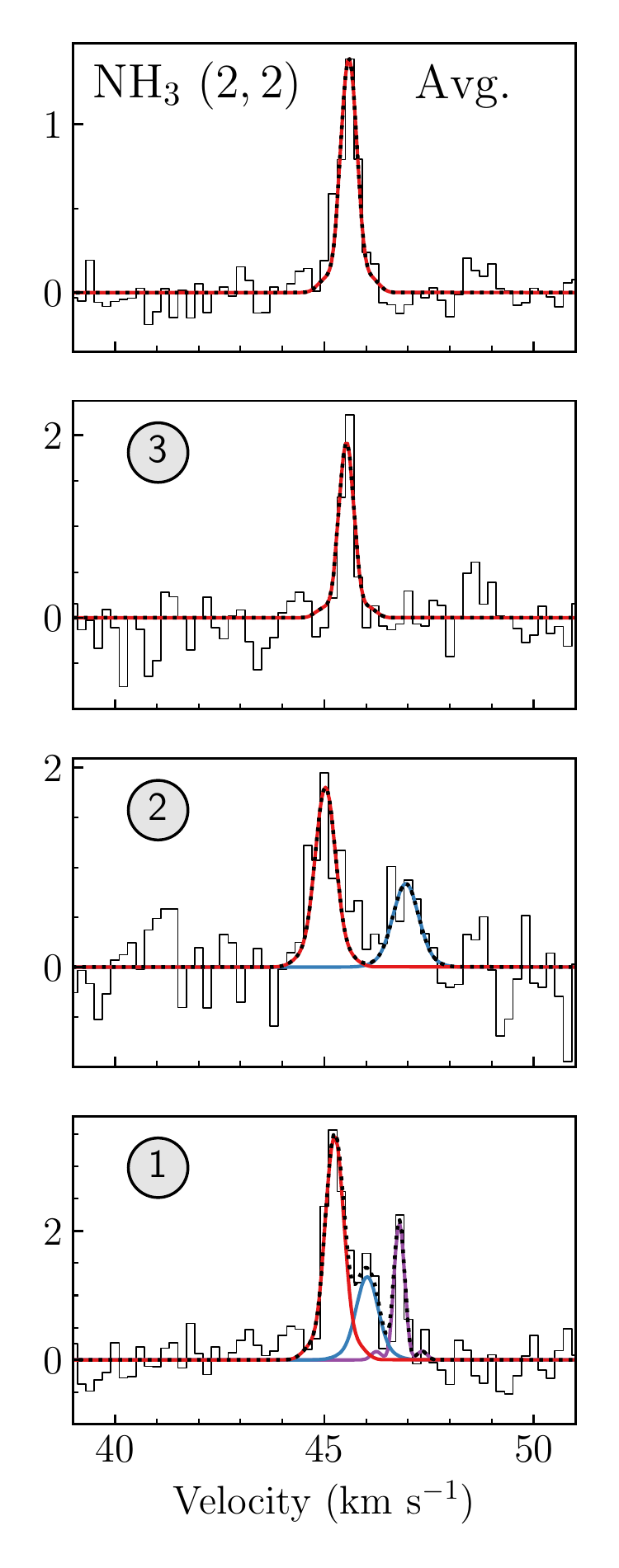}
    \caption{A map of the non-thermal velocity dispersion to the gas sound speed in G035.39. In pixels where multiple velocity components are discovered, the smallest value is shown on the image (the analyses throughout the rest of this work use all the values). The markers are the same as in Fig. \ref{fig:overview}, and the solid white line contours indicate the transition at $\sigma_{\rm nt} / c_{\rm s} = 1$. \hl{The side panels to the left and right of the map show (1,1) and (2,2) spectra \added{towards the numbered positions}. The spectra are overlaid with their best-fit model, with individual velocity components plotted in colour. The topmost spectra, unlike the others, show an averaged spectrum across the subsonic island at $\mathrm{\alpha(J2000) = 18h57m08s}$, $\mathrm{\delta(J2000) = +2\degr09\arcmin45\arcsec}$. An independently conducted fit, shown in red, yields $\mathcal{M} = 0.72 \pm 0.04$.}}
    \label{fig:mmap}
\end{figure*}

% forward-fitting is a thing! seriously, look it up and check that NASA link
The two ammonia spectral cubes were simultaneously forward-fitted pixel by pixel with the ammonia spectral profile model using the \textsc{pyspeckit} \citep{pyspeckit} Python package and the \nhhh formalism described in Friesen, Pineda et al. (\citeyear{friesen+pineda+2017}). This produced a set of one-, two- and three-component fits to each position,\footnote{When inspected, no individual spectrum appeared to contain more that three distinct velocity components.} where every fit has optimised values of the excitation temperature, kinetic gas temperature, total ammonia column density, velocity dispersion, and the velocity centroid of the line. Subsequently we make a heuristical decision on the number of velocity components in each spectrum \hl{(Fig. \ref{fig:mmap} shows selected spectra)} by using the same approach as in \cite{sokolov+2017}, namely, by restricting the minimal velocity separation between the components (components' FWHM are not allowed to overlap), and imposing a minimal peak signal to noise ratio ($\mathrm{S/N} > 3$) for each component.

% The one-figure-to-rule them all! Saves space, but the spectra are not as large as they could be in their own figure

In order to quantify the degree of non-thermal motions in the cloud, we calculate the non-thermal line widths by subtracting in quadrature both the thermal line widths and the channel width from the observed line widths.
% $\Delta v_{nt}^2 = \sqrt{\Delta v_{obs}^2 - \Delta v_{therm}^2 - \Delta v_{chan}^2}$.
\added{The thermal line widths are determined for each component using the fitted kinetic temperatures, which have a mean of 12.0 K and a standard deviation of 2.2 K across the entire map.}
We then derive the ratio of the non-thermal velocity dispersion component to the sound speed in the medium, $\mathcal{M} \equiv \sigma_{\rm nt} / c_{\rm s}$ for all the fitted components.
\hl{We propagate the uncertainties on the gas kinetic temperature and the line widths into the Mach number uncertainties.
The 25th, 50th, and 75th percentiles for the uncertainty distributions} are $\sigma_{T_\mathrm{d}}= $ 1.7, 2.5, and 3.0 K, respectively, for the gas temperatures, and 
$\sigma_{\mathcal{M}} =$ 0.079, 0.120, and 0.192, respectively, for the Mach number errors.

We present the spatial distribution of the Mach number, $\mathcal{M}$, in Fig. \ref{fig:mmap}. In case of more than one velocity component determined along the line of sight (\added{${\sim}7$\%}), the one with the smallest $\mathcal{M}$ is shown. While this choice was made for visual clarity, we emphasise that in all other analyses in this paper the full range of the $\mathcal{M}$ values is used. The map shown in Fig. \ref{fig:mmap} reveals the presence of multiple regions of subsonic motions in the G035.39, \hl{hereafter referred to as islands of coherence.}

More than a third (38.8\% of 7680) of all the spectral components have non-thermal gas motions in the subsonic regime ($\mathcal{M} < 1$), while 42.0\% are mildly supersonic ($1 < \mathcal{M} < 2$). The remaining data, ${\sim}19.2$\% of positions, show the non-thermal gas motions higher than the sound speed ($\mathcal{M} > 2$).
In addition to their spatially clumped appearance,
the subsonic \hl{islands} exhibit a higher degree of coherence. \hl{The standard deviation of the transonic and supersonic line width population is} 0.15 and 0.40 \kms respectively, consistently higher than the 0.08 \kms computed on the subsonic lines of ammonia. %This is reminiscent to the coherent core regions in low-mass star-forming regions.

\section{Discussion} \label{sec:discussion}

\begin{figure}
    \centering
    \includegraphics[width=0.52\textwidth]{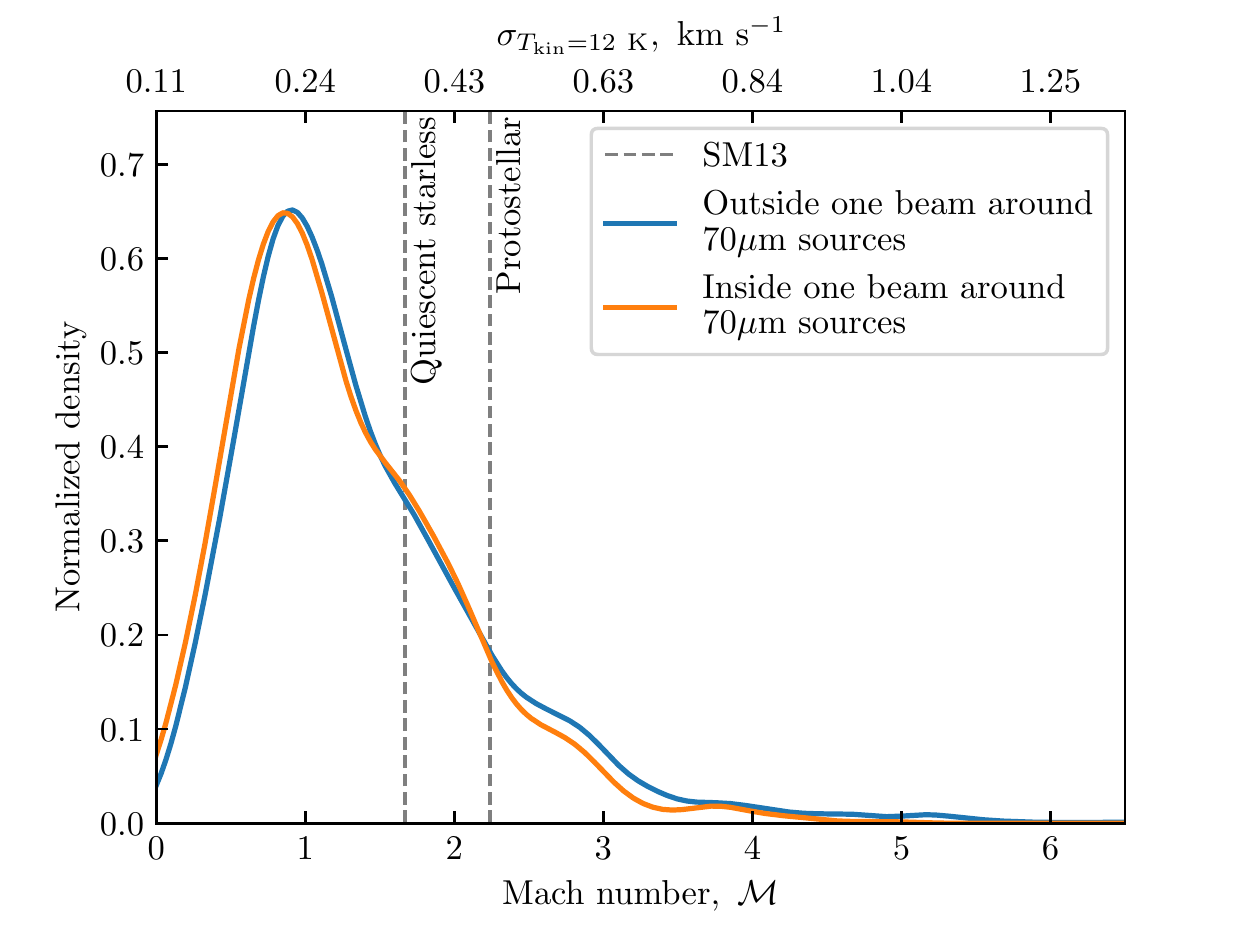}
    \caption{\hl{KDE distribution} of the non-thermal line widths to sound speed ratios in G035.39, peaking at $\mathcal{M} = 0.91$. The upper axis shows equivalent velocity dispersions for $T_{\rm kin} = 12$ K, a mean temperature in our sample.
    A subsample extracted from pixels within one synthesized beam of the VLA around the 70\um Herschel sources \citep{nguyen_luong+2011} is shown in orange, alongside the remaining data plotted in blue. Overplotted for the reference are the mean values for starless and protostellar IRDC cores found in \cite{sanchez-monge+2013}.}
    \label{fig:hist}
\end{figure}

Regions of massive star formation are generally thought to harbour internal motions larger than those of their lower-mass counterparts. To illustrate the difference between the previously found degrees of non-thermal motions and the ones found in G035.39, we plot \hl{the Gaussian kernel density estimate (KDE) of the} sonic Mach number distribution for all the fitted ammonia components in Fig. \ref{fig:hist}.
In a VLA survey of 15 IRDCs, \cite{sanchez-monge+2013} find that in 79 cores of differing evolutionary stages, the earliest ``starless core'' class has typical line widths and temperatures of 1.0 \kms and 16 K. While these values are considerably lower than the corresponding ones for the protostellar stages studied (cf. Fig. \ref{fig:hist} for the average values of \signt), they are still above the typical values found in G035.39 (\citealt{henshaw+2014} find mean $\mathcal{M}$ of 1.4--1.6 for the three filaments towards the northern part of the IRDC, and a fraction, $21\%$, of their spectra is subsonic).
A similar VLA survey of IRDC kinematics by \cite{ragan+2012} found no values of $\mathcal{M}$ below 2, favouring an even higher degree of non-thermal motions present inside the massive dense cores.
Three other VLA surveys of ammonia (\citealt{lu+2014}; upper limit of 1.3 \kms on line widths in \citealt{bihr+2015}; typical $\mathcal{M}$ from 3 to 5 reported in \citealt{dirienzo+2015}) report highly supersonic gas motions within IRDCs.
We suggest that the coarser spectral resolution of the previous VLA ammonia studies in similar IRDCs might have inhibited the detection of the subsonic regime, as our spectral resolution of 0.2 \kms is higher than that of the previous studies
(0.6 \kms in \citealt{ragan+2012, lu+2014, dirienzo+2015}; 0.6 \kms for most of the data used by \citealt{sanchez-monge+2013}; 0.8 \kms in \citealt{bihr+2015}).
Our results thus point to consistently lower line widths and temperatures than those found in other IRDCs, and closer to the typical values in nearby Gould Belt low-mass star-forming regions
(e.g. Friesen, Pineda et al. \citeyear{friesen+pineda+2017}).

% biases discussion - the instrumental contribution - even with the 0.2 km/s resolution we reach subsonic regime.
It is plausible to assume that the velocity dispersions we find might also be suffering from an instrumental bias due to insufficient spectral resolution. There are two reasons why this does not diminish our findings. Firstly, the mean observed line width of 0.71 \kms is well resolved with the channel width of 0.2 \kms, and the ammonia emission, split among many hyperfine transitions, has its spectral line profile described across a large number of channels. Therefore, constraining the velocity dispersions is typically not a problem (mean uncertainty on velocity dispersions: 0.015 \kms). Secondly, even if there is a spectral resolution bias present, it will bias the obtained velocity dispersions towards larger values \citep{friesen+2009}, thus only strengthening our claim of the subsonic motion regime detection. 
We note that even if our analyses are repeated without subtracting the channel width from the observed line width, we still recover a significant fraction of the subsonic spectra (0.36 of the all the fitted components vs. the original 0.39 fraction).
Additionally, the distribution of the line widths we obtain might be biased due to misidentification of the number of velocity components along the line of sight. 
As our heuristics to choose the number of components depend on all components in both ammonia lines being significantly detected, we expect a fraction of the multiple-component spectra to be fitted with a broader component model when those conditions are not met.
This misspecification \added{would only produce broader, not narrower, line widths, and may explain some Mach numbers in the tail of the Fig. \ref{fig:hist} distribution, as well as a sharp, border-like transition in the northern part of the IRDC (Fig. \ref{fig:mmap}). Despite this bias, 87\% of the subsonic components belong to spectra with single velocity component.}

Figure \ref{fig:hist} shows a subsample of values derived within one FWHM of the synthesized VLA beam around the protostellar 70\um sources identified in \cite{nguyen_luong+2011} alongside the remaining data.
As seen from the figure, the overall distribution of the non-thermal motions around the protostellar sources does not show a clear deviation away from that of the rest of the IRDC, indicating 
that the embedded sources do not yet exert enough feedback on the surrounding material to disturb the cores that harbour them (at least not on the \hl{0.07 pc} scales resolved with our VLA observations).
\hl{On the other hand, neither do the star-forming cores appear to resemble the "coherent core" picture of low-mass star formation (no correlation of $\mathcal{M}$ with $A_{\rm V}$; Pearson's $r = 0.10$).
It is likely that to disentangle these effects, a higher-angular-resolution kinematics study is needed to resolve the densest gas structures, similar in size to the narrow (0.03 pc) filaments found in a recent ALMA continuum image of G035.39 \citep{henshaw+2017_alma}.}
\hl{The importance of high-angular-resolution observations is also shown by \cite{hacar+2018}, who have recently resolved the Orion integral shaped filament into narrow 0.035 pc fibers of mostly subsonic nature, and a recent high-spectral-resolution study of SDC13 that shows localized traces of subsonic motions in the combined JVLA and GBT \nhhh observations \citep{williams+2018}.}
\hl{While future higher-angular-resolution studies dedicated to IRDC kinematics are needed in order to resolve the scales of the massive cores within the IRDCs, and probe the exact nature of transition into subsonic regime, our results show that some of the star-forming cores are forming in a quiescent environment.}

The narrow ammonia line widths found in G035.39 indicate that
\added{due to reduced turbulent support parts of the IRDC are prone to gravitational collapse, unless they are supported by magnetic fields.}
\cite{henshaw+2016_pdbi} find that magnetic field strengths of a few hundred microGauss are needed to virialize dense cores in the northern part of the IRDC, while
\cite{tan+2013} have observed four massive clumps in a sample of ten well studied IRDCs from \cite{butler+tan2009, butler+tan2012}, finding that magnetic fields of up to 1 mG strength are needed to support the cores in virial equilibrium \citep[cf. ${\sim}2$ mG in a follow-up study of][]{kong+2017}.
Likewise, \cite{zhang+2015} analysed thermal dynamic properties of dense cores in IRDC G028.34, and found that a magnetic field strength of several milliGauss is required in order to virialize these cores.
While the subsonic regions identified in our VLA observations do not always correspond to the density structure and therefore are not exclusively tracing the star-forming cores, we argue that by considering the largest bound $\mathcal{M} < 1$ contours to be magnetized cores in virial equilibrium we can arrive at an order of magnitude estimate of the magnetic field strength needed to support the enclosed region against collapse.
For the three largest continuous subsonic \hl{islands} in G035.39, each spanning in excess of seven VLA beam areas, we estimate masses of 22--45 M$_{\odot}$ from the mass surface density map of \cite{kainulainen+tan2013} (or 18--21 M$_{\odot}$ if the smallest value of mass surface density is representative of the line-of-sight contribution). For these values, assuming spherical core geometry, we follow the approach of \cite{tan+2013} to derive the magnetic field strengths of ${\sim}1.5\mbox{--}2~\mathrm{mG}$ (${\sim}0.8\mbox{--}1.8~\mathrm{mG}$ for the background subtraction case) needed to virialize the three regions.

Polarization studies of massive star-forming cores find that magnetic fields play an important role during their collapse and fragmentation \citep{zhang+2014}, and field strengths up to a few milliGauss are supporting active high-mass star-forming regions \citep{frau+2014, qiu+2014, li+2015, pillai+2016}.
Recent polarization measurements towards early-stage IRDCs \citep{pillai+2015, santos+2016, beuther+2018} find field strengths of a few hundred microGauss to a few milliGauss, that is, the same order as our estimates for G035.39.
\added{As G035.39 hosts a number of protostellar sources, it is clear that parts of the IRDC are already unstable or undergoing gravitational collapse.}
Dust polarization observations of G035.39, together with resolved
\added{kinematics tracing accreting and infalling motions of the dense gas,} will reveal a comprehensive picture on the stability of this IRDC.

\hl{Our findings, enabled by high spectral resolution of the combined VLA and GBT observations, allow us to quantify the islands of subsonic turbulence within the IRDCs for the first time. The results of this work indicate that early stages of massive star and cluster formation can go through stages more similar to their low-mass counterparts than previously thought.
We highlight the potential for high-angular- and spectral resolution ALMA observations of IRDC kinematics, needed to reevaluate the role of turbulent dissipation and investigate the exact nature of transition to coherence in early stages of massive star- and cluster-forming regions.
If the subsonic regions found here are representative of the kinematics of the massive dense star-forming cores in other IRDCs, the reduced turbulent support and possible increased support from the magnetic fields put constraints on models and simulations aiming to reproduce the initial stages of massive star and cluster formation.}

\begin{acknowledgements}
VS, JEP, and PC acknowledge the support from the European Research Council (ERC; project PALs 320620).
KW is supported by grant WA3628-1/1 of the German Research Foundation (DFG) through the priority program 1573 ("Physics of the Interstellar Medium").
JCT acknowledges NASA grant 14-ADAP14-0135.
IJ-S acknowledges the financial support received from the STFC through an Ernest Rutherford Fellowship (proposal number ST/L004801/2).
This research made use of Astropy, a community-developed core Python package for Astronomy \citep{astropy}, and of APLpy, an open-source plotting package for Python \citep{aplpy}.

\end{acknowledgements}

\bibliography{subsonic-irdc}

\end{document}